\documentclass[aps,prd,preprint,superscriptaddress,showpacs,showkeys]{revtex4}
\usepackage{amssymb,amsmath}
\usepackage{graphicx} 
\usepackage{dcolumn}  
\usepackage{epsfig}   
\usepackage{pstricks}
\usepackage{ifpdf}
\newcommand{\Tr}{ {\mathrm{Tr}\, }}

\begin{document}


\title{On the absence of the Gribov copy problem in \emph{effective locality} QCD calculations }


%
%

\author{T. Grandou}
\affiliation{Universit\'{e} cote d'Azur\\ Institut de Physique de Nice, UMR CNRS 7010; 1361 routes des Lucioles, 06560 Valbonne, France}
\email[]{Thierry.Grandou@inphyni.cnrs.fr}
\author{R. Hofmann}
\affiliation{Institut f\"ur Theoretische Physik\\ 
Universit\"at Heidelberg\\ 
Philosophenweg 16\\ 
69120 HEIDELBERG}
\email[]{r.hofmann@thphys.uni-heidelberg.de}



\date{\today}

\begin{abstract} About ten years ago the use of standard functional manipulations was demonstrated to imply an unexpected property satisfied by the fermionic Green's functions of QCD and dubbed \textit{effective locality}. This feature of QCD is non-perturbative as it results from a full \emph{gauge invariant} summation of the gluonic degrees of freedom. This astounding result has lead to suspect that in a way or other, the famous Gribov copy problem had been somewhat overlooked. It is argued that it is not so. 

 \end{abstract}


\maketitle

\section{\label{SEC:1}Introduction}

In some recent articles~\cite{QCD1,QCD-II,QCD5, QCD6, QCD5'} a property of the non-perturbative fermionic Green's functions of QCD was put forward under the name of {\textit{effective locality}} and summarized as follows. 
\par
\textit{For any fermionic $2n$-point Green's functions and related amplitudes, the
full gauge-fixed sum of cubic and quartic gluonic interactions,
fermionic loops included, results in a local contact-type interaction. This local interaction is
mediated by a tensorial field which is antisymmetric both in Lorentz and
color indices. Moreover, the resulting sum appears to be fully gauge-fixing independent, that is, gauge-invariant.}
\par
 This is surprising because integrations of elementary degrees of freedom ordinarily result in non-local 
 structures, \cite{WilsonLoop1974}, and the `effective locality' denomination has been coined to account for this unusual circumstance. Notice that in the pure euclidean Yang Mills case, and up to the first non-trivial orders of a semi-classical expansion, the same phenomenon of effective locality was observed a welcome property in an attempt to construct a formulation \emph{dual} to the original Yang Mills theory \cite{RefF}. Now, apart from a \textit{supersymmetric} extension with $\mathcal{N}=2$, \cite{SW1994}, $QCD$ is not known to possess any dual formulation, and the full effective locality functional expressions certainly attest to this situation. However, like in the pure Yang Mills case of \cite{RefF}, effective locality is a useful means to learn about non-perturbative physics in $QCD$, and has the appreciable advantage of  proceeding from first principles. 

\par\medskip\medskip
 In a previous paper entitled \emph{Effective Locality and the non-perturbative realisation of QCD gauge-invariance} the full derivation of effective locality was revisited displaying it as a robust property of non-perturbative $QCD$ \cite{fgh}. In particular it was proven that effective locality holds independent of any choice of a representation used for $G_F(x,y|A)$ and $L[A]$, the quark fields propagator in a background gauge field configuration $A$, and closed quark loop functionals, respectively (See (\ref{Z}) and (\ref{Fradkin}) below). 
\par
Even more important was the fact that gauge invariance appeared guaranteed as a matter of gauge fixing independence, and that the famous issue of \emph{Gribov copies} did not seem to plague effective locality calculations. This non-trivial statement certainly requires substantiation and is the main concern of this letter. 

\par
The paper is organized as follows. The next Section is a reminder of effective locality and is made as concise as possible relying on the example of a $4$-point fermionic Green's function. Further, some approximations known to preserve both effective locality and Gribov's problematics will be used with the double purpose of simplifying the current presentation while reaching sound conclusions. The matter of \emph{Gribov copies} is examined in Section III, as well as the possible \emph{non-perturbative reach} of effective locality calculations. Considerations on the equivalence of gauge invariance and gauge-fixing independence conclude this Section, and an overall brief conclusion is proposed in Section IV.

 \section{\label{SEC:2}A reminder on effective locality}
 \subsection{Effective locality in short}
 Let us begin with a reminder of what effective locality is. 
\noindent Effective locality surfaces at the level of fermionic Green's functions once functional differentiations of the following generating functional with respect to the sources $\bar{\eta}$, $\eta$ are taken and are subsequently set to zero, $\bar{\eta}=\eta=j_\mu^a=0$, \cite{fgh},
   \begin{eqnarray}\label{Z}
{Z}_{\mathrm{QCD}}[j, \bar{\eta}, \eta] &=& \mathcal{N}e^{\frac{i}{2} \int{j \cdot {D}_{\mathrm{F}}^{(0)} \cdot j}}\nonumber\\ &\times& \left. e^{- \frac{i}{2} \int{\frac{\delta}{\delta A} \cdot {D}_{\mathrm{F}}^{(0)} \cdot \frac{\delta}{\delta A} } } \cdot e^{-\frac{i}{4} \int{{F}^{2}} + \frac{i}{2} \int{ A \cdot \left( -\partial^{2}\right) \cdot A} } \cdot e^{i\int{\bar{\eta} \cdot {G}_{\mathrm{F}}[A] \cdot \eta} + {L}[A]}\right|_{A = \int{{D}_{\mathrm{F}}^{(0)} \cdot j} }
\end{eqnarray}where ${{D}_{\mathrm{F}}^{(0)}}_{\mu\nu}=g_{\mu\nu}D_F$ is the Feynman propagator ($\partial^2D_F=\delta^4$). The fermionic functional $G_F(x,y|A)$ can be represented with the help of a \emph{Fradkin representation} such as \cite{RefL},
 \begin{eqnarray}\label{Fradkin}
&& {{G}^{\,\alpha\beta}_{\mathrm{F\,ij}}}_{}(x,y|A) = i \int_{0}^{\infty}{ds \ e^{-is m^{2}}} \, e^{- \frac{1}{2} \Tr{\ln{\left( 2h \right)}} } \, \int{d[u]} \, e^{ \frac{i}{4} \int_{0}^{s}{ds' \, [u'(s')]^{2} } } \, \delta^{(4)}(x - y + u(s))\nonumber \\  & & \quad\quad\quad\quad\quad\quad\!\!\times {\left[ m\delta^{\alpha\beta}\delta_{ik} + i g \gamma^{\alpha\beta}_{\mu} A^{\mu}_{a}(y-u(s)) \lambda^{a}_{ik} \right]} \nonumber \\  & & \quad\quad\quad\quad\quad\quad\!\!\times  \left( T_{s'}e^{ -ig \int_{0}^{s}{ds' \, u'_{\mu}(s') \, A^{\mu}_{a}(y-u(s')) \, \lambda^{a}} + g \int_{0}^{s}{ds' \sigma^{\mu \nu} \, {F}_{\mu \nu}^{a}(y-u(s')) \, \lambda^{a}}} \right)_{kj}
\end{eqnarray}with $\lambda^as$ the $SU_c(3)$-Lie-algebra generators taken in the fundamental representation and where the $\sigma_{\mu\nu}=[\gamma_\mu,\gamma_\nu]$ are the usual Lorentz group generators. Indices $\alpha, \beta$ are spinorial and $i,j=1,2,3$ are colour indices. One has also,
\begin{equation} h(s_1,s_2)=s_1\Theta(s_2-s_1)+s_2\Theta(s_1-s_2)\ ,\ \ \ h^{-1}(s_1,s_2)=\frac{\partial}{\partial s_{1}} \frac{\partial}{\partial s_{2}} \delta(s_1-s_2).\end{equation}The fermionic closed loop functional $L[A]=\Tr \ln i{\rlap /\!{D}}(A)$, with $i\rlap /\!{D}(A)$ the Dirac operator, can be endowed with a Fradkin representation similar to the latter \cite{QCD1,QCD-II,QCD5, QCD6, QCD5'}. In (\ref{Fradkin}), the 4-vector $u(s)$ is the Fradkin variable, while in the last line the $T_{s'}$- indicates an $s'$-Schwinger-proper-time ordering of the expression between parenthesis.

\noindent  In order to derive the property of effective locality, one introduces a `linearization' of the $\int F^2$ term which appears in the right hand sides of (\ref{Z}). This can be achieved by using the representation, \cite{RefF,Halpern1977a,Halpern1977b},
\begin{equation}\label{Eq:17}
e^{-\frac{i}{4} \int{{F}^{2}}} = \mathcal{N}' \, \int{\mathrm{d}[\chi] \, e^{ \frac{i}{4} \int{ \left(\chi_{\mu \nu}^{a}\right)^{2} + \frac{i}{2} \int{ \chi^{\mu \nu}_{a} {F}_{\mu \nu}^{a}} } } }
\end{equation}
\noindent where, 
\begin{equation}\int{\mathrm{d}[\chi]} = \prod_{z} \prod_{a} \prod_{\mu <\nu} \int{\mathrm{d}[\chi_{\mu \nu}^{a}](z)}\,.
\end{equation} As usual, space-time is broken up into small cells of infinitesimal size $\delta^{4}$ about each point $z$, and $\mathcal{N}'$ is a normalization constant so chosen that the right hand side of (\ref{Eq:17}) becomes equal to unity as ${F}_{\mu \nu}^{a} \rightarrow 0$.  In this way, the generating functional (\ref{Z}) may be re-written as (with $\mathcal{N}' \cdot \mathcal{N} \equiv \mathcal{N}''\rightarrow \mathcal{N}$),
\begin{eqnarray}\label{Z1}
{Z}_{\mathrm{QCD}}[j,\bar{\eta},\eta] = \mathcal{N} \, \int{\mathrm{d}[\chi] \, e^{ \frac{i}{4} \int{ \chi^{2} }} } \, \left. e^{\mathfrak{D}_{A}^{(0)}} \cdot e^{-\frac{i}{2} \int{\chi \cdot {F} + \frac{i}{2} \int{ A \cdot \left( -\partial^{2}\right) \cdot A} }} \cdot e^{i\int{\bar{\eta} \cdot {G}_{\mathrm{F}}[A] \cdot \eta} + {L}[A]}\right|_{A = \int{{D}_{\mathrm{F}}^{(0)} \cdot j} }
\end{eqnarray}
\noindent where the shorthand notation is introduced,  
\begin{equation}\label{link}\mathfrak{D}_{A}^{(0)} = - \frac{i}{2} \int \mathrm{d}^4x\int \mathrm{d}^4y\ {\frac{\delta}{\delta A^a_\mu(x)} \left. {D}_{\mathrm{F}}^{(0)}\right|^{ab}_{\mu\nu}(x-y)\, \frac{\delta}{\delta A_\nu^b(y)}}\,,\end{equation} and where the integration on $\chi$ is permuted with the action of the functional differentiations operated by the so-called \emph{linkage operator}, $\exp\,\mathfrak{D}_{A}^{(0)}$.
\par\medskip
Of course, using the representation (\ref{Eq:17}) breaks the manifest gauge invariance of the left hand side. This can be remedied in several ways. For example, one can gauge the $\chi^a_{\mu\nu}$-fields, imposing them to transform in exactly the same way as the field-strength tensors, so that the term $(\chi\cdot F)$ recovers a manifest invariance. In \cite{RefF}, this route was followed successfully in order to reach a (semi-classical) formulation dual to the pure euclidean Yang Mills theory. Proceeding in this way however, the gauge fixing originally bearing on the potentials $A^a_\mu$ gets transferred to some $\chi$-field functional, namely, $\mathcal{J}^a_\mu(z)=[(\partial_\lambda\chi^{\lambda\nu}_b(z))[f\cdot\chi)^{-1}(z)]^{ba}_{\nu\mu}$, and so generates also a Gribov-copy problem in the final $\chi$-field integration.
\par
There exists another possibility. If it is possible to complete the $\chi^a_{\mu\nu}$-functional integrations in an exact way, then, one is guaranteed to deal with the full invariant left hand side of (\ref{Eq:17}). At some mild \emph{eikonal} approximation at least, and in the strong coupling limit this is doable by using a standard analytic continuation of the powerful \emph{Random Matrix} calculus \cite{QCD6,RefI}. It should be noted that the strong coupling condition, $g>>1$, is in no way restrictive for our purpose, because it is in this very limit that \emph{the Gribov ambiguity} is precisely known to come about \cite{Singer,IZ}. Likewise, in order to simplify the analysis, the closed quark loop functional $L[A]$ can be ignored since it is, in itself, a gauge-invariant functional \cite{Herb}.
\par
\par
For the sake of an alleviate presentation, it is convenient to proceed with a simplified derivation of effective locality.

\subsection{Effective locality at eikonal approximation}

There is no loss of generality illustrating things on the basis of a 4-point fermionic Green's function as, through more cumbersome expressions, the following structures generalise easily to the case of $2n$-point fermionic Green's functions (see \cite{RefI}, Appendix D). Then, two propagators $G_F(x_1,y_1|A)$ and $G_F(x_2,y_2|A)$ are to be represented with the help of  (\ref{Fradkin}). Upon substitution into (\ref{Z1}), a representation like (\ref{Fradkin}) produces a cumbersome structure of an exponential of an exponential. This is why it is necessary to bring (\ref{Fradkin}) down by means of functional differentiations with respect to the \emph{Grassmannian sources} $\bar{\eta},\eta$, and in this way one deals with $2n$-point fermionic Green's functions. Accordingly, in contrast to the field strength formulation dual to the pure Yang Mills case \cite{RefF}, the property of effective locality is referred to the behavior of fermionic Green's functions rather than to that of the generating functional itself. Of course, this is not a restriction of generality because fermionic Green's functions exhaust the whole fermionic content of the theory.
\par\medskip
In (\ref{Fradkin}) though, the linear and quadratic $A^a_\mu$-field dependences are contained within a time-ordered exponential which prevents the linkage operator (\ref{link}) to operate in a simple way. This can be circumvented at the expense of introducing two extra field variables so as to remove the $A^a_\mu$-field variables from the ordered exponential. For example, 
\begin{equation}\label{out}
T_s\left(e^{ig\,p^\mu\!\int_{-\infty}^{+\infty} {\rm{d}}s\,A^a_\mu(y-sp)\,T^a}\right)=\mathcal{N}\int {\rm{d}}[\alpha]\int{\rm{d}}[\Omega]\, e^{-i\!\int_{-\infty}^{+\infty} {\rm{d}}s\,\,\Omega^a(s)[\alpha^a(s)-gp^\mu A^a_\mu(y-sp)]}T_s\left(e^{i\int_{-\infty}^{+\infty}{\rm{d}}s\,\alpha^a(s)T^a}\right)
\end{equation}where $\mathcal{N}$ is a normalisation constant. Equation (\ref{out}) defines an \emph{eikonal} approximation of (\ref{Fradkin}) which is used here to offer a simple derivation of the property of interest. 
\par
Now, not until the full integrations on the two auxiliary field variables $\alpha^a$ and $\Omega^a$ are brought to completion, can one be assured to deal with the proper Fradkin representation of $G_F[A]$. At this level of approximations though and in the strong coupling limit, $g\gg 1$, it is a fortunate circumstance that these extra dependences can be integrated out in an exact way thanks to an analytically continued \emph{Random Matrix} calculus \cite{{QCD6},{RefI}}.
\par
Skipping over the details of integrations on Schwinger proper times, Fradkin's variables and the extra fields $\alpha^a_i$ and $\Omega^a_i$, $i=1,2$, one obtains a result of the following form
\begin{eqnarray}\label{ELEQ}
\prod_{i=1}^{2}\int\mathrm{d}s_i \int\mathrm{d}u_i(s_i)\int\mathrm{d}\alpha_i(s_i)\int\mathrm{d}\Omega_i(s_i)\,\left(\ddots\right) \int{\mathrm{d}[\chi] \, e^{ \frac{i}{4} \int{ \chi^{2} }} } \, \left. e^{\mathfrak{D}_{A}^{(0)}} \, e^{{+ \frac{i}{2} \int{ A ^a_\mu\, K^{\mu\nu}_{ab}\, A^b_\nu} }} \, e^{i\int{Q^a_\mu A^\mu_a } }\right|_{\ A \rightarrow 0 }\,,
\end{eqnarray}
where $K^{\mu\nu}_{ab}$ and $Q^a_\mu$ \emph{factorise} the quadratic and linear $A^a_\mu$-field dependences, respectively,
\begin{equation}\label{QK}
K_{\mu\nu}^{ab}=gf^{abc}\chi_{\mu\nu}^c+\left({{D}_{\mathrm{F}}^{(0)}}^{-1}\right)_{\mu \nu}^{a b}\,, \ \ \ Q^a_\mu =-\partial^\nu\chi^a_{\mu\nu}+g[R^a_{1,\mu}+R^a_{2,\mu}]\,, \ \  \ f^{abc}\chi^c_{\mu\nu}=
(f\cdot \chi)^{ab}_{\mu\nu}\,,\end{equation}and where the $R^a_{i,\mu}$ arise from the part of (\ref{out}) which is linear in the potential fields $A^a_\mu$,
 \begin{equation}\label{currents}
 R^a_{i,\mu}(z) = p_{i,\mu}\, \int\mathrm{d}s_i\,\Omega^a_{i}(s_i)\,\delta^4(z - y_{i} + s_i\, p_{i})\,, \ \ \ \ \ i=1,2\,. 
 \end{equation}
  Note that in (\ref{currents}) the eikonal approximation has substituted $s_ip_i$ for the 
 original Fradkin field variable $u_i(s_i)$ ({\textit{i.e.}} a straight line approximation, connecting the points $z$ and $y_i$).
The linkage operation followed by the prescription of setting the sources $j^a_\mu$s equal to zero is now trivial and yields,
\begin{eqnarray}\label{elform}
& & \left. e^{-\frac{i}{2} \int{\frac{\delta}{\delta A} \cdot {D}_{\mathrm{F}}^{(0)} \cdot  \frac{\delta}{\delta A} }} \cdot e^{+ \frac{i}{2} \int{A \cdot {K} \cdot A} + i \int{A \cdot {Q} }}  \right|_{A \rightarrow 0} \\ \nonumber &=& e^{-\frac{1}{2} \Tr{\ln{\left( 1- {D}_{\mathrm{F}}^{(0)} \cdot {K} \right)}}} \cdot e^{\frac{i}{2} \int{{Q} \cdot \left[ {D}_{\mathrm{F}}^{(0)} \cdot \left( 1 - {K} \cdot {D}_{\mathrm{F}}^{(0)}\right)^{\!-\!1} \right]\cdot {Q}}}.
\end{eqnarray}

\noindent On the right hand side, the kernel of the quadratic term in ${Q}_{\mu}^{a}$ is
\begin{eqnarray}\label{magics}
{D}_{\mathrm{F}}^{(0)} \cdot \left( 1 - {K} \cdot {D}_{\mathrm{F}}^{(0)}\right)^{\!-\!1} &=& {D}_{\mathrm{F}}^{(0)} \cdot \left( 1 - \left[ g f \cdot \chi + {{D}_{\mathrm{F}}^{(0)}}^{\!-\!1} \right] \cdot {D}_{\mathrm{F}}^{(0)}\right)^{\!-\!1} \\ \nonumber &=& - \left( g f \cdot \chi \right)^{\!-\!1},
\end{eqnarray}so that eventually,
\begin{eqnarray}\label{EL}
& & \left. e^{-\frac{i}{2} \int{\frac{\delta}{\delta A} \cdot {D}_{\mathrm{F}}^{(0)} \cdot  \frac{\delta}{\delta A} }} \cdot e^{+ \frac{i}{2} \int{A \cdot {K} \cdot A} + i \int{A \cdot {Q} }}  \right|_{A \rightarrow 0} \\ \nonumber &=& e^{-\frac{1}{2} \Tr{\ln{\bigl[-g{D}_{\mathrm{F}}^{(0)}\bigr]}}} \cdot \frac{1}{\sqrt{\det(f\cdot\chi)}}\cdot e^{-\frac{i}{2} \int\mathrm{d}^4z\ {{Q}(z) \cdot (gf\cdot\chi(z))^{-1}\cdot {Q}(z)}}\,,
\end{eqnarray}where the first term is a constant (possibly infinite) which can be absorbed into a redefinition of the overall normalization constant $\mathcal{N}$. 
\par
The manifestation of effective locality is in the last term of (\ref{EL}). While the $\left({\mathbf{D}_{\mathrm{F}}^{(0)}}\right)_{\mu \nu}^{a b}$-pieces entering (\ref{elform}) and (\ref{magics}) are \emph{non-local}, they disappear from the final result so as to leave the \emph{local} structure $[gf\cdot\chi]^{-1}$ which represents an effective local  interaction between quarks,
\begin{equation}\label{local}
\langle x | (gf\cdot\chi)^{-1} | y \rangle = (gf\cdot\chi)^{-1}(x) \, \delta^{(4)}(x-y)\,.
\end{equation} This offers a simple way to look at the effective locality property whose detailed derivation relies in an essential way on the non-abelian character of the gauge group \cite{fgh}. Contrary to expectations in effect \cite{RefO}, this phenomenon cannot show up in the abelian case of QED~\cite{fgh}. Furthermore, thanks to another remarkable consequence of effective locality, the final integration on the $\chi^a_{\mu\nu}$-fields lends itself to a(n analytically continued \cite{QCD6}) Random Matrix exact calculation \cite{{QCD6},{RefI}}.  
\par
Returning to (\ref{ELEQ}), one finds a result whose final form reads as, 
\begin{eqnarray}\label{ELEQ1}
\mathcal{N}\,\prod_{i=1}^{2}\int\mathrm{d}s_i \int\mathrm{d}u_i(s_i)\,\delta^{(4)}(x_i-y_i+u_i(s_i))\int\mathrm{d}\alpha_i(s_i)\int\mathrm{d}\Omega_i(s_i)\,\left(\ddots\right)  \nonumber\\ \cdot\,{\left[ m + \gamma_\mu\lambda^a\frac{\delta}{\delta R^a_{i,\mu}(x)} \right]} \int{\mathrm{d}[\chi] \, e^{ \frac{i}{4} \int{ \chi^{2} }} }\frac{1}{\sqrt{\det(f\cdot\chi)}}\cdot e^{-\frac{i}{2} \int\mathrm{d}^4z\ {{Q}(z) \cdot (gf\cdot\chi(z))^{-1}\cdot {Q}(z)}}\,,\end{eqnarray}
where nothing ever more refers to ${\mathbf{D}_{\mathrm{F}}^{(0)}}$, that is to any choice of a gauge fixing condition chosen so as to make of $[(\partial^2)(g^{\mu\nu}-\partial^\mu\partial^\nu/\partial^2)]^{-1}$ a well defined propagator. 

\par\medskip
 As demonstrated in \cite{fgh}, in effect, had one relied on any other \emph{covariant}, \emph{non-covariant}, \emph{axial-planar}, \emph{Fock-Schwinger} choice of  a gauge-fixing condition, non-linear gauge conditions included, exactly the same result as (\ref{ELEQ1}) was obtained. This is gauge-invariance as a matter of full gauge-fixing independence as will be discussed in next Section. 
 \par
 Actually, as recalled in \cite{Thess}, to calculate observables it is not mandatory to fix a gauge, and no \emph{real} gauge-fixing is ever used in the derivation of effective locality. This is certainly unusual a way to proceed, but not exceptional either \cite{Greensite}. \par
 Moreover, the limitations of a gauge-fixing procedure should be kept in mind in view of the unsolved Gribov problem \cite{Thess} of non-perturbative $QCD$, while, \emph{even} in the perturbative sector the gauge-fixing procedure (\emph{unitary} vs.$\,R_\xi$) is being questioned for perturbative calculations in the {Standard Model} Higgs-particle decay into two phtons, mediated through a $W$ loop \cite{TaiWu}.

 \section{On the absence of Gribov copies}
 \subsection{Effective locality and Wick theorem}
 In the strong coupling/field regime, it has been known for long that the problem of Gribov copies compromises a fully controlled gauge invariance of $QCD$ observables as computed by non-perturbative approaches to the respective functional integrals over gauge field configurations. In addition, effective locality provides an approach to the non-perturbative regime of $QCD$ where the Gribov problem does not seem to show up. 
 \par
 Concisely stated, there is no Gribov copy problem in an effective locality calculation, because in the absence of any \emph{real} gauge-fixing procedure there is no gauge-fixing condition to be possibly copied \cite{fgh}. For the sake of accuracy, it is worth recalling that by \emph{real}, the following is meant. An arbitrary gauge-fixing density is added to \emph{and} subtracted from the full $QCD$ lagrangian density in order to make sense of some intermediate operation \cite{QCD1,fgh}. Now, the gauge-fixing density being added \emph{and} subtracted, no gauge-fixing condition is really implemented in an effective locality calculation. Accordingly, results come out independent of any sort of gauge-fixing condition, as they should.
 \par\medskip\noindent
 This astounding fact, concerning the non-perturbative regime of $QCD$, needs to be explained in more details.
 \par
 The key-point is poorly known. It is that quantising a field theory can be achieved either by functional integrations on $c$-fields, but equivalently, by functional differentiations with respect to the same fields, followed by a prescription of setting the sources of these fields to zero in calculating Green's functions. This equivalence, whose extent will be discussed shortly, can be viewed as the equivalence of two different expressions of the standard \emph{Wick's theorem} for vacuum averages of time-ordered products of field operators.
 \par
 One of these two equivalent forms is expressed in terms of functional integrations, as is textbook material \cite{RefO1, IZ1, Sterman}, and is most commonly used for the sake of symmetry considerations. Not so customary in the literature, is the expression in terms of functional differentiation. Besides \cite{RefL} and \cite{Herb}, a noticeable and recent exception is that of \cite{Kleinert} in which a functional-differentiation form of Wick's theorem is explicitly demonstrated in full details in the case of a simple scalar field theory. For the sake of completeness some of these steps are summarised now. 
 \par
 With $T$ and $N$ the usual prescriptions for \emph{Time} and \emph{Normal orderings}, one can prove the basic relation
 \begin{equation}\label{basic}
 T\left(\varphi(x_1)\varphi(x_2)\right)=N\left(\varphi(x_1)\varphi(x_2)\right)+\langle T\,\varphi(x_1)\varphi(x_2)\rangle\end{equation}whose last term is the Feynman propagator $G_F(x_1,x_2)\equiv \langle T\varphi(x_1)\varphi(x_2)\rangle$. By induction (\ref{basic}) extends to an arbitrary number of field operators so that the original/operatorial form of Wick's theorem can be given as the following identity,
  \begin{equation}\label{Wick}
 T(\varphi(x_1)\dots \varphi(x_n))=\sum_{all\,pairs\,of\,T-\,contractions}N(\varphi(x_1)\dots \varphi(x_n))\end{equation}where an example of a \emph{T-contraction} is given by $G_F(x_1,x_2)$, the last term of (\ref{basic}). Now, the familiar functional differentiation identity, $\delta\varphi(x)/\delta\varphi(x')=\delta^{(4)}(x-x')$, allows to rewrite (\ref{basic}) as,
  \begin{equation}\label{basic2}
 T\left(\varphi(x_1)\varphi(x_2)\right)=\left(1- \frac{1}{2}\int\mathrm{d}^4y\int\mathrm{d}^4z\,\frac{\delta}{\delta\varphi(y)}\,G_F(y,z)\,\frac{\delta}{\delta\varphi(y)}\right)\,N\left(\varphi(x_1)\varphi(x_2)\right)\end{equation}where fields are treated as $c$-fields when acted upon with the functional differentiation, and where the beginning of the linkage operator expansion, (\ref{link}), can be noticed. A few more standard steps are to be taken so as to provide Wick's theorem with the functional differentiation form,
 \begin{equation}\label{fnal}
 T\left(e^{i\int j\varphi}\right)= e^{-\frac{1}{2}\int\mathrm{d}^4y\int\mathrm{d}^4z\,\frac{\delta}{\delta\varphi(y)}\,G_F(y,z)\,\frac{\delta}{\delta\varphi(y)}}\,N\left(e^{i\int j\varphi}\right)\end{equation}
where the extension from polynomials of field operators, as in (\ref{basic2}), to exponentials of field operators holds true. In (\ref{fnal}), the functional differentiation operator of (\ref{link}) can be fully recognised, passing of course from scalar, $\varphi$, to non-abelian gauge fields, $A^a_\mu$.
\par\medskip
Now, the following point is \emph{crucial} to any effective locality calculation and cannot be overemphasized. Relation (\ref{fnal}) is but a pure mathematical identity, relying on well defined mathematical steps, each. The functional differentiation operator of (\ref{fnal}) just operates the summation $\sum_{all\,pairs\,of\,T-\,contractions}$ of (\ref{Wick}). It is accordingly self contained and needs not being referred to any functional integration process from which (\ref{fnal}) would be derived. 
 \par\medskip
 Before proceeding with the limits of the equivalence alluded to above, one may quote another form of it, derived in a formal way involving $c$-fields instead of field operators. This can be seen as follows. Let $\mathcal{F}[A]$ be a functional of a $c$-field $A$, and evaluate $ \left. e^\mathfrak{D}\,\mathcal{F}[A]\right|_{A =0}$, with $\exp\mathfrak{D}$ the linkage operator with $\mathfrak{D}$ given in (\ref{link}). Omitting integrations on spacetime for short, one can write,
 \begin{eqnarray}\label{rodos}
 \left. e^\mathfrak{D}\,\mathcal{F}[A]\right|_{A =0}&=&\frac{\int\mathrm{d}[B]\,e^{\frac{i}{2}(B-D\frac{\delta}{\delta A})D^{-1}(B-D\frac{\delta}{\delta A})}\, \left. e^\mathfrak{D}\,\mathcal{F}[A]\right|_{A =0}}{\int\mathrm{d}[B]\,e^{\frac{i}{2}(B-D\frac{\delta}{\delta A})D^{-1}(B-D\frac{\delta}{\delta A})}}\nonumber \\ && \!\!\!\!\!\!\!=\frac{\int\mathrm{d}[B]\,e^{\frac{i}{2}BD^{-1}B}\,e^{-iB{\delta\over \delta A}}\,\left.\mathcal{F}[A]\right|_{A =0}
 }{\int\mathrm{d}[B]\,e^{\frac{i}{2}BD^{-1}B}}\nonumber \\ && \!\!\!\!\! \!\!=\frac{\int\mathrm{d}[B]\,e^{\frac{i}{2}BD^{-1}B}\,\mathcal{F}[B]}{\int\mathrm{d}[B]\,e^{\frac{i}{2}BD^{-1}B}}\,,\end{eqnarray}an identity which, again, holds for polynomial as well as exponential functionals, $\mathcal{F}[A]$ \cite{RefL}. 
  \par\medskip\noindent
 The point here is that a closer inspection at (\ref{rodos}) quickly reveals that  (\ref{rodos}) doesn't hold in the general case. While the left hand side is mathematically well defined, and this even beyond the purpose of defining an \emph{asymptotic series} in the coupling constant \cite{timoutcha} (when $\mathcal{F}[A]=\mathcal{F}[g;A]$ is expandable as a power series in $g$), the right hand side is not. This problem arises because the integration measure $\mathrm{d}[B]$  usually is not defined. Only in a few cases the measure $\mathrm{d}[B]$ can be endowed with a sound mathematical definition \cite{timoutcha,Ted}. Besides, not even demanding the level of mathematical rigour available in \emph{Wiener functional spaces} \cite{Ted}, in a $QCD$-$BRST$ quantisation procedure the Gribov problem prevents $\mathrm{d}[B]$ to be defined in a controlled enough way, beyond perturbation theory \cite{IZ,Lowdon}. 
 
    \par\medskip\noindent  
 This is why, from a mathematical point of view, one often takes things the other way round, the left hand side of (\ref{rodos}) being fluently proposed as a definition of the right hand side to which it can also provide analytic continuation to the space of non-singular quadratic forms \cite{ timoutcha}; and one writes \cite{IZ, timoutcha}, 
  \begin{equation}\label{Def}
\frac{\int\mathrm{d}[B]\,e^{\frac{i}{2}BD^{-1}B}\,\mathcal{F}[B]}{\int\mathrm{d}[B]\,e^{\frac{i}{2}BD^{-1}B}}\,\textbf{:=}\,  \left. e^\mathfrak{D}\,\mathcal{F}[A]\right|_{A =0}\,. \end{equation} 
  
  \par\medskip\noindent
 Now, for our current concern it is of utmost importance to realise that it is in the left hand side way of (\ref{rodos}) that the property of effective locality is derived in the simplified (eikonal and quenched) case of the current letter, as (\ref{Z}), (\ref{Z1}), (\ref{ELEQ}), (\ref{elform}) and (\ref{ELEQ1}) amply testify, as well as in the full $QCD$ case \cite{QCD-II}. That is, \emph{without} any gauge-fixing, \emph{nor} any subsequent ill-defined measure of integration on a functional space of connexions $A^a_\mu$ for which it is recognised that there exists no functional coordinate system able to cover the whole of it \cite{Singer}.
 \par\medskip
 This allows one to appreciate the considerable simplification brought about by effective locality calculations as they enjoy two most important features. 
 \begin{itemize}
 \item They proceed along mathematically well defined steps. 
 \item They avoid definitely the intractable Gribov copy problem inherited from gauge-fixing \emph{and} functional integration in the strong coupling/field limit.\end{itemize}
But `how much non-perturbative' is an effective locality calculation?
 \par\medskip
   \subsection{On the non-perturbative reach of effective locality calculations}
 The following considerations go beyond a simple matter of terminology. As stated in the Introduction in effect, the property of effective locality is said to be \emph{non-perturbative}, and has always been presented as such \cite{QCD1,QCD-II,QCD5, fgh, RefI,tgpt}. On the other hand it has just been argued that effective locality proceeds from a re-summed standard Wick theorem expansion, and therefore is rooted in perturbation theory.
 \par\medskip
  (i) For physicists, effective locality will easily be taken as a non-perturbative property. This is because the whole sum of gauge field mediated interactions between quarks is carried out, in a gauge-fixing independent way, achieved along the lines of functional differentiation, the left hand side of (\ref{rodos}). Whereas the whole Wick expansion of (\ref{elform}) formally is a series expansion in the coupling constant $g$, its summation (\ref{magics}) generates an order $g^{-1}$ non-perturbative output, a result which extends to the full (non-approximate) $QCD$ theory~\cite{QCD-II}.
  \par
   It may be recalled here that in the pure euclidean Yang Mills case this transformation of $g\rightarrow g^{-1}$ is the fundamental \emph{criterion} of a duality relation to the original theory, satisfied at first non-trivial orders of a semi-classical expansion \cite{RefF}. As stated in the Introduction, though, effective locality does not furnish the bases of a plain duality relation. In particular it is easily seen from (\ref{EL}) that the basic dual relation of $g\rightarrow g^{-1}$ is not satisfied. Still, the `duality content' of effective locality allows for a description of at least a part of the non-perturbative regime of $QCD$. This goes for example for dynamical chiral symmetry breaking which cannot result from perturbation theory \cite{tgpt}.
  \par\medskip
  (ii) As stated in the analysis of \cite{timoutcha}, the mathematical point of view is different, as the right hand side of (\ref{Def}) would \emph{only} define a \emph{perturbative evaluation} of the path integral. There is no contradiction in these two different points of view. The mathematical point of view \cite{timoutcha} only presupposes that there could exist others, non-perturbative ways of evaluating the left hand side of (\ref{Def}) directly, while recognising that this is not achieved yet.
 \par
 But there is more to it, and it comes in support to the physical point of view (i). The mathematical theory of \emph{Resurgence} \cite{Ecalle} allows one to appreciate that a perturbative re-summed Wick expansion such as the right hand side of (\ref{Def}), really encodes non-perturbative physics, and \emph{Generating Non-perturbative Physics from Perturbation Theory} \cite{Dunne} is by now extensively studied \cite{Zinn}. This requires to extend to complex fields the original real ones \cite{Guralniks}, with functional measures of integration which are no better defined than in the latter case. Up to this proviso, however, the indication persists that re-summed perturbative expansions are not de-correlated from the non-perturbative content of a theory, see also \cite{Renormalons}. It is the merit of Resurgence theory to have provided sound mathematical bases to this long suspected connection \cite{Carl}.

  \subsection{Gauge invariance versus gauge-fixing independence}
  So far, as stated in the Introduction, gauge-fixing independence has always been taken as synonymous of gauge invariance. In the context of an exact Renormalisation Group analysis though, it has been noticed that the two notions need not necessarily correspond to each other \cite{Rosten}. It is this very point which is examined in this subsection.
  \par\medskip  1. That gauge invariance involves gauge-fixing independence is an obvious statement. In order to evaluate observables, physicists are free to use any gauge they find convenient, precisely because observables are gauge invariant. However circuitous and involved actual procedures may be in practice, in the cases of $QED$ and perturbative $QCD$, in the end, gauge-invariance implies/requires gauge-fixing independence.
  \par\medskip
  2. The converse is not easy to see because the check of gauge-fixing independence often relies on explicit perturbative expansions in an order-by-order analysis. It is obvious in the case of the \emph{partition function}, though, as it involves no functional differentiations with respect to the fermionic sources.

  \par\medskip
 Now, a $2n$-point fermionic Green's function is given by an expression,
\begin{equation}\label{2n}
X^{(2n)}(..,x_i,y_i,\dots)\equiv \mathcal{N} \left. e^{- \frac{i}{2} \int{\frac{\delta}{\delta A} \cdot {D}_{\mathrm{F}}^{(\zeta)} \cdot \frac{\delta}{\delta A} } } \cdot e^{-\frac{i}{4} \int{{F}^{2}} + \frac{i}{2} \int{ A \cdot \left({ {D}_{\mathrm{F}}^{(\zeta)}}\right)^{-1} \cdot\, A} } \,\, \prod_{i=1}^nG_F(x_i,y_i|A) \right|_{A =0 }\,,
\end{equation}where $F^2$ is the full non-abelian strength-field tensor, and $D_{\mathrm{F}}^{(\zeta)}$ some covariant free propagator \cite{fgh}, an arbirary intermediate step of an effective locality calculation and an element of ${\mathcal{D}}$, the \emph{generalised} functional space of all possible free propagators (propagators being \emph{distributions}). The simplification of quenching is used in (\ref{2n}), since the closed quark loop functional $L[A]$ has no bearing on the current considerations. \textit{A priori}, one has $X^{(2n)}=X^{(2n)}[D_{\mathrm{F}}^{(\zeta)}]$, a functional of $D_{\mathrm{F}}^{(\zeta)}$, as it happens to be in the abelian case of $QED$ for example \cite{fgh}.
\par
Out of $X^{(2n)}$, though, the following average can be defined, 
\begin{equation}\label{2nn}
Y^{(2n)}\equiv \mathcal{N} \frac{\int_{\mathcal{D}} \mathrm{d}[{D_{\mathrm{F}}^{}}^{ab}_{\mu\nu}]\left. e^{- \frac{i}{2} \int{\frac{\delta}{\delta A} \cdot {D_{\mathrm{F}}^{}} \cdot \frac{\delta}{\delta A} } } \cdot e^{-\frac{i}{4} \int{{F}^{2}} + \frac{i}{2} \int{ A \cdot \left({ {D_{\mathrm{F}}^{}}}\right)^{-1} \cdot A} }\, \prod_{i=1}^nG_F(x_i,y_i|A) \right|_{A =0 }}{\int_{\mathcal{D}} \mathrm{d}[{D_{\mathrm{F}}^{}}^{ab}_{\mu\nu}]}\,,
\end{equation}that is,
\begin{equation}\label{2nnn}
Y^{(2n)}(..,x_i,y_i,\dots)\equiv  \frac{\int_{\mathcal{D}} \mathrm{d}[{D_F}^{ab}_{\mu\nu}]\  X^{(2n)}(..,x_i,y_i,\dots)\,[D_F^{}]}{\int_{\mathcal{D}} \mathrm{d}[{D_F}^{ab}_{\mu\nu}]}\,.
\end{equation}In an effective locality calculation, at the level of $X^{(2n)}(..,x_i,y_i,\dots)[D_F^{}]$, the only gauge dependence which is retained is in the gauge propagator $D_F^{}$. Averaging over the full generalised functional space ${\mathcal{D}}$, a gauge-invariant result $Y^{(2n)}(..,x_i,y_i,\dots)$ is therefore obtained by construction.

\noindent Moreover, in (\ref{2nn}) and (\ref{2nnn}), in contradistinction to (\ref{rodos}) and (\ref{Def}) where the measure $\mathrm{d}[B]$ is not defined, the measure of integration $\mathrm{d}[D^{ab}_{\mu\nu}]$ can now be defined properly by resorting to the construction used already in \cite{QCD6, tgpt,RefI}, which amounts to transform the propagator $[D^{ab}_{\mu\nu}]$ into the sum of two real symmetric matrices of format $N\times N$, where $N=D(N_c^2-1)=32$, at $4$ spacetime dimensions and $N_c=3$ colours. With $1\leq i,j\leq N$ and $n=N_c^2-1=8$,
\begin{equation}\label{}
D^{ab}_{\mu\nu}\longrightarrow \mathbb{M}^{ij}+i\mathbb{N}^{ij}\,,\ \   i=a+\mu n\,,\,\,\,j=b+\nu n\,,\end{equation}where ${\mathbb{N}}$ accounts for the imaginary part of $D^{ab}_{\mu\nu}$. For any of the two matrices $\mathbb{K}=\{\mathbb{M},\mathbb{N}\}$, with $\xi_i(\mathbb{K})$ their eigenvalues, one has now \cite{Mehta,QCD6, tgpt, RefI},       
\begin{equation}\label{}
\int\mathrm{d}\mathbb{K}=\prod_{i=1}^N \int_{-\infty}^{+\infty}\mathrm{d}\xi_i(\mathbb{K})\,\,\prod_{l<k}^N |\,\xi_l(\mathbb{K})-\xi_k(\mathbb{K})|\,\, \int_{O_N(\mathbb{R})}\mathrm{d}\mathcal{O}(\mathbb{K})\end{equation}where the last factor stands for a \emph{Haar measure} of integration on $O_N(\mathbb{R})$, the group of orthogonal matrices with real coefficients. 
\par
This allows one to see in a \emph{definite} way that if $X^{(2n)}[D_F]$ is gauge-fixing independent, being independent of $D_F$, as effective locality establishes, (\ref{elform}) and (\ref{magics}), then the \emph{volume} of the functional $D_F$-space factorises in the numerator of (\ref{2nn}) and (\ref{2nnn}) (possibly requiring some regularisations) and one obtains eventually,
\begin{equation}\label{ouf}
X^{(2n)}[D_F]=Y^{(2n)}\,,\end{equation}that is, a proof that in the context of an effective locality calculation, gauge-fixing independence implies gauge-invariance. Not even evoking the problem of the measure, the same proof in the context of functional integration is certainly much harder to establish, supposing even that it is true, and falls beyond the scope of the current letter.
\par\medskip
To sum up, in the very context of effective locality, gauge-fixing independence and gauge-invariance should be taken as equivalent. 
 
\section{Conclusion}
That an approach to the non-perturbative regime of $QCD$ may avoid the 40 years old Gribov problem may sound somewhat preposterous. In this letter however it has been argued that this is what effective locality calculations actually accomplish. At least \emph{formally} in the case of non-approximate $QCD$, and \emph{concretely} in the case of a mild eikonal and quenching approximations, and in the strong coupling limit which is relevant to this issue. In this latter situation, this is because auxiliary field integrations can be carried out exactly, guaranteeing a full control of gauge invariance.
In a future publication it will be shown, in accord to the duality aspect of effective locality, that this property lends itself to sound perturbative developments for the originally non-perturbative sector of $QCD$.

 \end{document}